\documentclass[prc,twocolumn,showpacs,preprintnumbers,superscriptaddress,
               amsmath,amssymb]{revtex4}
\usepackage{graphicx}          
\usepackage{dcolumn}           
\usepackage{bm}                

\usepackage{color}
\usepackage{lipsum}
\usepackage{booktabs}
\usepackage{kantlipsum}
\usepackage{booktabs} 

\begin{document}

\preprint{\fbox{\sc version of \today}}

\title{Spectroscopic Criteria for Identification of Nuclear Tetrahedral and Octahedral Symmetries: Illustration on a Rare Earth Nucleus}
       
\author{J.~Dudek}
\affiliation{{\it Universit\'e de Strasbourg, CNRS, IPHC UMR 7178,
             F-67\,000 Strasbourg, France}\\
	         }
\affiliation{{\it Institute~of~Physics,~Marie Curie-Sk\l odowska University,
             PL-20\,031 Lublin, Poland}\\
	         }
\author{D.~Curien}
\affiliation{{\it Universit\'e de Strasbourg, CNRS, IPHC UMR 7178,
             F-67\,000 Strasbourg, France}\\
	         }

\date{\today}
\author{I.~Dedes}
\affiliation{{\it Institute~of~Physics,~Marie Curie-Sk\l odowska University,
             PL-20\,031 Lublin, Poland}\\
	         }
\author{K.~Mazurek}
\affiliation{The Niewodnicza\'nski Institute of Nuclear Physics, Polish Academy
             of Sciences, 2 ulica ~Radzikowskiego 152, PL-31\,342 Krak\'ow, Poland}
\author{S.~Tagami}
\affiliation{Department of Physics, Faculty of Sciences, Kyushu University,
             Fukuoka 8190359, Japan}
\author{Y.~R.~Shimizu}
\affiliation{Department of Physics, Faculty of Sciences, Kyushu University,
             Fukuoka 8190359, Japan}
\author{T.~Bhattacharjee}
\affiliation{Variable Energy Cyclotron Centre, IN-700\,064 Kolkata, India}

\date{\today}
 
\begin{abstract}
We formulate criteria for identification of the nuclear tetrahedral and octahedral symmetries and illustrate for the first time their possible realization in a Rare Earth nucleus $^{152}$Sm. We use realistic nuclear mean-field theory calculations with the phenomenological macroscopic-microscopic method, the Gogny-Hartree-Fock-Bogoliubov approach and the general point-group theory considerations to guide the experimental identification method as illustrated on published experimental data. Following group-theory the examined symmetries imply the existence of exotic rotational bands on whose properties the spectroscopic identification criteria are based. These bands may contain simultaneously states of even and odd spins, of both parities and parity doublets at well defined spins. In the exact-symmetry limit those bands involve no E2-transitions. We show that coexistence of tetrahedral and octahedral deformations is essential when calculating the corresponding energy minima and surrounding barriers and that it has a characteristic impact on the rotational bands. The symmetries in question imply the existence of long-lived shape-isomers and, possibly, new waiting point nuclei -- impacting the nucleosynthesis processes in astrophysics -- and an existence of 16-fold degenerate particle-hole excitations. Specifically designed experiments which aim at strengthening the identification arguments are briefly discussed.
\end{abstract} 


\maketitle

%
%
%

This article presents and illustrates a technique of identifying the so-called high-rank symmetries -- tetrahedral and/or octahedral ones -- in non $\alpha$-cluster nuclei. The technique is based on combining group theory and realistic nuclear energy calculation results, whereas illustration is based on published experimental data. These symmetries have so far not been identified in the present context and we believe that the proposed approach provides a powerful tool in their research. 

A characteristic impact of the high-rank symmetries on the studies of nuclear stability was discussed in the general mean-field context in Ref.\,\cite{JDu10}. Tetrahedral symmetry was predicted in Ref.\,\cite{JDu02} to exist in nuclei in various areas of the Periodic Table, following earlier predictions in Ref.\,\cite{XLi94} and followed further in Ref.\,\cite{JDu06}. The chains of {\em tetrahedral magic numbers} -- $Z/N=16,20,32,40,56,64,70,90$ and $N=112,136,142$ -- which correspond to increased nuclear stability at the tetrahedral geometry have been proposed in Ref.\,\cite{JDu02}. A partial review of the evolution in the studies addressing various theoretical aspects of tetrahedral symmetry in sub-atomic physics can be found in Ref.\,\cite{JDu13}. A coexistence between octupole (in particular tetrahedral) geometries in selected light nuclei has been studied in Refs.\,\cite{STa98,MYa01}, see also \cite{KZb06,KZb09,JZh17,POl06}. Discussion of tetrahedral symmetry effects in very heavy nuclei can be found in Refs.\,\cite{PJa11,YSC08,YSC10,PJa17}. A recent interpretation of the spectrum of $^{16}$O in terms of the 4-$\alpha$ cluster-structure and tetrahedral symmetry, based on the algebraic models, can be found in Ref.\,\cite{RBi14}, see also references therein.

Below we examine the co-existence and inter-relations between the two symmetries. We demonstrate on the example of $^{152}$Sm, that using {\em simultaneously} what we call tetrahedral and octahedral deformations is essential when studying the implied exotic-symmetry minima. Therefore certain earlier calculations using the tetrahedral deformations alone may need to be revisited.

Despite of the fact that tetrahedral and octahedral symmetries are well known in molecular physics, their identification in heavy nuclear systems such as Rare Earths or Actinide nuclei remained elusive. One of the difficulties consisted in the fact that the quadrupole and dipole moments in nuclei with the exact high-rank symmetries vanish. It then follows that the electric quadrupole (E2) and/or electric dipole (E1) transitions -- usually dominating in the feeding and in the decay paths -- in the case of high-rank symmetries are simply absent. Thus, even though a tetrahedral/octahedral-symmetric nucleus is expected to generate rotational bands, neither E2-, nor E1 intra-band transitions can be used as signals which help identify these structures. In what follows we will show how to transform this instrumental difficulty into a positive argument serving our method of analysis and helping the symmetry-identification.

It is instructive to begin by recalling certain elementary mathematical properties of high-rank groups and their irreducible representations. Recall that the tetrahedron is invariant under 24 symmetry-elements such as plane reflections and rotations through specific angles about 2, 3, and 4-fold axes. All these operations form a tetrahedral group, $T_d$. However, to describe symmetries of fermion Hamiltonians, a related, so-called double point group is needed. The corresponding tetrahedral double ($T^D_d$) group is composed of 48 elements and characterized by the presence of two 2-dimensional and one 4-dimensional irreducible representations. The reader unfamiliar with the group theory notions will merely need to remember that the single-nucleon levels in the mean-field Hamiltonians symmetric with respect to the $T^D_d$ group form three families, two of them composed of levels obeying the usual two-fold degeneracy (sometimes referred to as spin-down spin-up or Kramers degeneracy) and one family of 4-fold degeneracy -- thus exotic levels. This latter property is unusual for atomic nuclei: it has never been identified in sub-atomic physics. 

The octahedral-symmetry double point group, $O^D_h$, is composed of 96 symmetry elements and characterized by four 2-dimensional and two 4-dimensional irreducible representations. This implies that the nucleonic energy levels in the octahedral-symmetric mean-field form four families of 2-fold and two families of 4-fold degenerate energy levels. Let us emphasize that the {\em tetrahedral group is a sub-group of the octahedral one} and, consequently, from the formal point of view, each octahedrally-symmetric object is at the same time tetrahedrally-symmetric. 


We refer to the shape of a nuclear surface, $\Sigma$, using the expansion in terms of the spherical-harmonics 
$\{Y_{\lambda\mu}(\vartheta,\varphi)\}$ and deformation parameters 
$\alpha_{\lambda \mu}$:
\begin{equation}
   \Sigma:\;
   R(\vartheta,\varphi)
   =
   R_0 c(\alpha)
   \bigg[ 1 + \sum_{\lambda=2}^{\lambda_{max}}
              \sum_{\mu=-\lambda}^{\lambda}
              \alpha^\star_{\lambda\mu} 
              Y^{}_{\lambda\mu}(\vartheta,\varphi)\bigg].
                                                                 \label{eqn.01}
\end{equation}
Above, $c(\alpha)$ assures that the nuclear volume is independent of the deformation, whereas symbol $\alpha$ represents the ensemble of all the deformation parameters $\{\alpha_{\lambda \mu}\}$. We have $R_0=r_0\,A^{1/3}$, where $r_0\approx 1.2$\,fm is the nuclear radius parameter. It has been demonstrated in Ref.\,\cite{JDu07}, that for $0 < \lambda \leq 9$, the only possible multipole expansions for the surfaces with tetrahedral symmetry are given by: 
\begin{eqnarray}
      t_1 
      &\equiv& 
      \alpha_{3,\pm2}, 
                                                                 \label{eqn.02}
                                                                 \\
      t_2 
      &\equiv& 
      \alpha_{7,\pm2} \;\; \textrm{and} \;\;
      \alpha_{7,\pm6} = -\sqrt{{11}/{13}}\, \alpha_{7,\pm2},
                                                                 \label{eqn.03}
                                                                 \\
      t_3 
      &\equiv& 
      \alpha_{9,\pm2} \;\; \textrm{and} \;\;
      \alpha_{9,\pm6} = +\sqrt{{13}/{3}}\;\;\, \alpha_{9,\pm2}.
                                                                 \label{eqn.04}
\end{eqnarray}
The above notation implies that, whereas tetrahedral-symmetric surfaces of the lowest order are expressed with the help of two spherical harmonics, 
$(Y_{3+2}+Y_{3-2})$, the next orders are characterized by two pairs of terms; for instance in the 7th order, $(Y_{7+2}+Y_{7-2})$ and $(Y_{7+6}+Y_{7-6})$. Moreover, as it happens, each order (here: $\lambda=3,7,9$) depends on {\em one single real parameter only}, which can be chosen as $t_1\equiv\alpha_{3+2}=\alpha_{3-2}$ and similarly  $t_2\equiv\alpha_{7+2}=\alpha_{7-2}$ and
$t_3\equiv\alpha_{9+2}=\alpha_{9-2}$.

Following Ref.\,\cite{JDu07} for the modeling of the {\em octahedral} symmetry shapes with $0<\lambda \leq 9$, the only solutions are:
\begin{eqnarray}
      o_1 &\equiv& \alpha_{4,0}
      \;\; \textrm{and} \;\;
      \alpha_{4,\pm 4}= -\sqrt{{5}/{14}}\,\alpha_{4,0},
                                                                 \label{eqn.05}
                                                                 \\
      o_2 &\equiv& \alpha_{6,0}
      \;\; \textrm{and} \;\;
      \alpha_{6,\pm4} = +\sqrt{{7}/{2}}\;\;\alpha_{6,0},
                                                                 \label{eqn.06}
                                                                 \\
      o_3 &\equiv& \alpha_{8,0}
      \;\; \textrm{and} \;\; 
      \alpha_{8,\pm4} = -\sqrt{{28}/{198}}\,\alpha_{8,0},
                                                                 \nonumber
                                                                 \\
      & & \quad\quad\; \textrm{and} \;\;
      \alpha_{8,\pm8} = +\sqrt{{65}/{198}}\,\alpha_{8,0}.
                                                                 \label{eqn.07}
\end{eqnarray}
Each octahedral-deformation depends on one real independent parameter with a possible choice: $o_1\equiv\alpha_{40}$,  $o_2\equiv\alpha_{60}$ and $o_3\equiv\alpha_{80}$.


Let us emphasize that the presence of the 4-fold degeneracy of certain nucleonic levels for both of these symmetries plays a central role in stabilizing the underlying shell-gaps and implied high-rank symmetry minima, as indicated already by other authors, see Fig.\,2 and surrounding text in Ref.\,\cite{JDu09} or Fig.\,1 and surrounding text in Ref.\,\cite{JDu10}. It implies in principle e.g.~a presence of the unprecedented 16-fold degenerate 1p-1h excitation energies as a possible sign of high-rank symmetries.

The remaining part of this presentation is focused on the nucleus $^{152}_{\;\;62}$Sm$^{}_{90}$, one of the four nearest neighbors of the tetrahedral doubly-magic $^{154}_{\;\;64}$Gd$^{}_{90}$. This choice is supported by the fact that for this particular nucleus there exist experimental data compatible with the group-theory requirements which very strongly constrain the excited spectrum and the decay properties. We have verified that in none of the nuclei in the neighborhood the analogous data exist. This may be partly due to the experimental limitations. Indeed, in the rare earth region, $^{152}$Sm is very special since about 25 nuclear reactions lead to its production. In surrounding nuclei, typically of the order of a dozen but usually much fewer reactions can be used. This limits the possibilities in neighboring nuclei to populate the levels of interest, which due to the vanishing E1 and E2 transitions, are fed via nuclear rather than electromagnetic transitions.

\begin{figure}[h!]
\begin{center}
\includegraphics[width=0.50\textwidth]{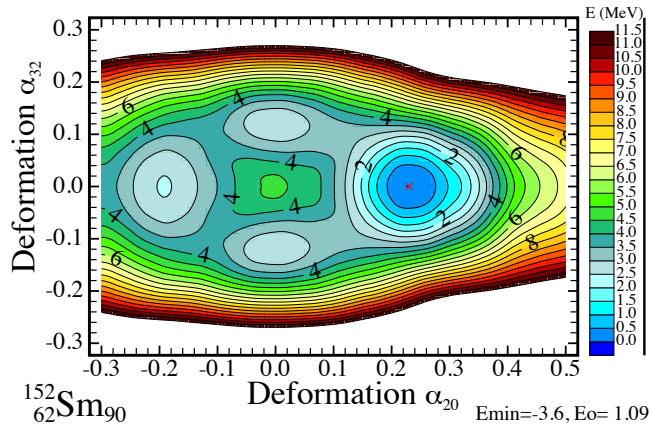}
\end{center}
\caption{Total nuclear energy for the nucleus $^{152}$Sm, calculated using the standard {\em macroscopic microscopic} method with the realistic phenomenological mean-field in the form of the so-called Woods-Saxon Universal realization of Ref.\,\cite{DDC78}. The vertical axis corresponds to the $\alpha_{32}$-multipole deformation which represents the tetrahedral-symmetry nuclear-shape. At each $(\alpha_{32},\alpha_{20})$-point the energy has been minimized over the octahedral deformations $o_1$ and $o_2$.  Here and in other Figures the energies are normalized assuming that the liquid-drop contribution is zero at zero deformation; $E_o$ gives then the shell and pairing energy at the spherical shape, whereas E$_{\rm min}$ gives the energy at the absolute minimum.}
                                                                 \label{fig.01}
\end{figure}

\begin{figure}[h!]
\begin{center}
\includegraphics[width=0.50\textwidth]{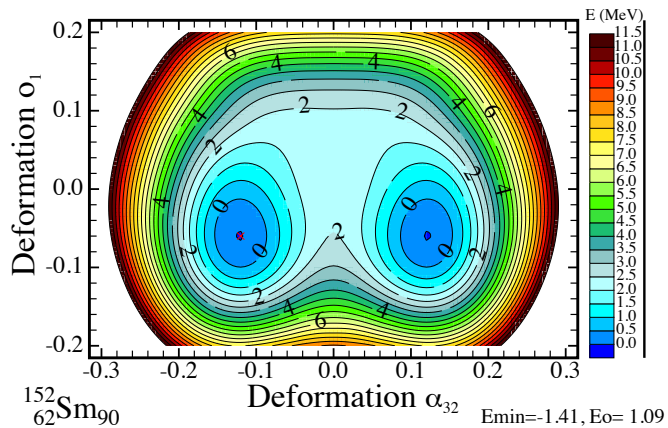}
\end{center}
\caption{Total energy surface of the $^{152}$Sm nucleus projected on the $(o_1,\alpha_{32})$-plane, $o_1$ defined in Eq.\,(\ref{eqn.05}). Here all other deformations are set to zero. }
                                                                 \label{fig.02}
\end{figure}

Realistic total energy calculations have been performed using the well tested phenomenological Woods-Saxon mean-field Hamiltonian of 
Ref.\,\cite{DDC78}. Figure \ref{fig.01} shows total energy projection on the $(\alpha_{32},\alpha_{20})$-plane for $^{152}$Sm. A pair of tetrahedral symmetry minima at $(\alpha_{20},\alpha_{32})=(0,\pm 0.12)$ deserves noticing. Figure \ref{fig.02} illustrates the impact of the {\em combined} tetrahedral and octahedral deformations. This combination brings the energy of the exotic minimum about 2.5\,MeV lower compared with the spherical configuration; nearly 40\% of this effect is due to the octahedral component. This property allows us to interpret the exotic-symmetry content in $^{152}$Sm, as a combined effect of the T$^D_d$ and O$^D_d$ symmetry groups.


\begin{figure}[ht!]
\begin{center}
\includegraphics[width=0.50\textwidth]{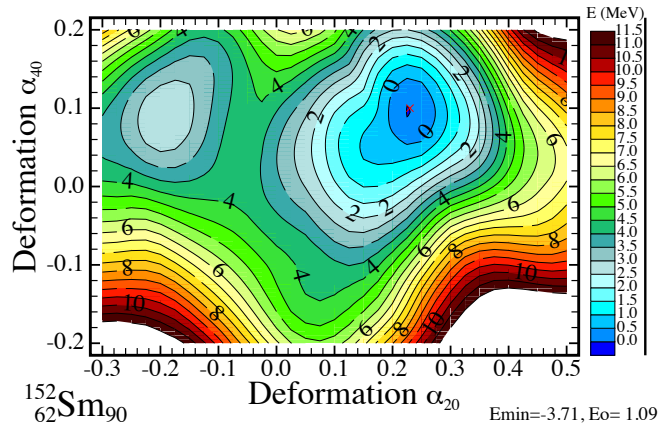}
\end{center}
\caption{Similar to Fig.\,\ref{fig.01} but for the axial-hexadecapole deformation $\alpha_{40}$ replacing the tetrahedral one, $\alpha_{32}$, at the vertical axis. For increasing quadrupole deformation an increase in the axial hexadecapole deformation causes a decrease in the energy of the ground-state minimum stronger as compared to the octahedral one, E$_{\rm min}$=-3.71 MeV compared to -2.60 MeV, cf.\,Figs.\,\ref{fig.01} and \ref{fig.04} (for explanation see text).}
                                                                 \label{fig.03}
\end{figure}


\begin{figure}[ht!]
\begin{center}
\includegraphics[width=0.50\textwidth]{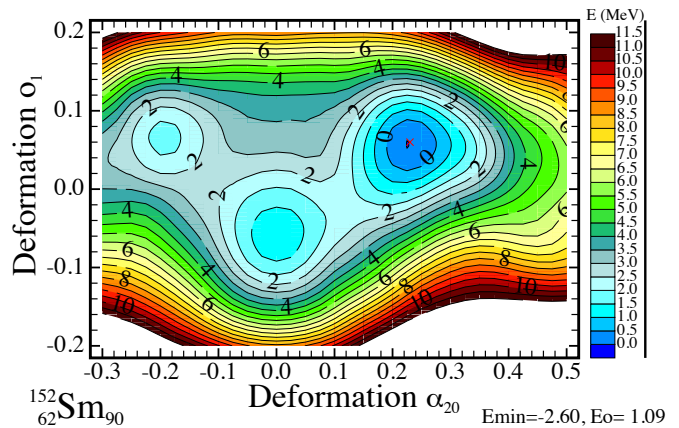}
\end{center}
\caption{Similar to Fig.\,\ref{fig.03}, but illustrating the impact of $o_1$. The absolute minimum lies higher, $E_{\rm min.}=-2.60$\,MeV compared to $E_{\rm min.}=-3.71$\,MeV in Fig.\,\ref{fig.03}.}
                                                                 \label{fig.04}
\end{figure}

Let us mention a technical issue of a simultaneous description of the  hexadecapole, $\alpha_{40}$, and the octahedral shape components, $o_1\propto\alpha_{40}$. Since octahedral deformation, Eq.\,(\ref{eqn.05}), involves a fixed proportion of $\alpha_{44}\neq 0$, formally, the hexadecapole and octahedral variables cannot be treated simultaneously (independently) and thus their effects have been studied separately. The presence of the  $\alpha_{44}$-component increases the effect of the surface energy in the macroscopic-microscopic energy expression compared to a pure $\alpha_{40}$ case. It follows that at the ground-state deformation the implied difference amounts to about 1.1\,MeV in favor of the pure axial hexadecapole deformation, as it can be seen from comparison of Figs.\,\ref{fig.03} and \ref{fig.04}. However, at $\alpha_{20}\approx 0$, the effect of the octahedral deformation is decisive in producing well pronounced high-rank symmetry total energy minima. Indeed, from Fig.\,\ref{fig.03} it can be seen that no local energy minimum is predicted at $\alpha_{20}\approx 0$, whereas Fig.\,\ref{fig.04} shows a well pronounced minimum at $\alpha_{20}\approx 0$.

To address the issue of an experimental identification of the lowest (thus in principle preferentially populated) rotational band with tetrahedral and/or octahedral shape-symmetries in $^{152}$Sm, let us note that the lowest-state in such a band is characterized by $I^\pi=0^+$. It then follows that the lowest energy exotic-symmetry rotational-band should contain this $I^\pi=0^+$ state. This represents an extra experimental difficulty since such a state should be populated directly via nuclear reaction and remains isomeric. The realistic lifetime estimates and form of decay of such states are out of reach for the present day theory. The metastable states of this type might decay {\em via} internal conversion or $\beta$-decay and be detected in specifically designed experiments. Alternatively one can envisage E3-transitions from the excited $3^-$ with decay-strengths 8-9 orders of magnitude weaker than the E2 transitions usually measured. As discussed below, we could find no evidence for this state in the existing data; however, neither could we find in the literature any reports about the specific experiments needed to detect the presence of this special state.

Tetrahedral group has five irreducible representations denoted in the literature $A_1$, $A_2$, $E$, $F_1$ and $F_2$. However, the  $0^+$ state, belongs uniquely to $A_1$, and so do all the other states of the "tetrahedral ground-state band". Inspecting standard decomposition tables, e.g.~Table VI of Ref.\,\cite{STa13}, allows us to construct the spin-parity sequence forming the {\em tetrahedral} $A_1$ band in question (we arbitrarily limit the discussion to states with spins $0 \leq I \leq 11$):
\begin{equation}
   A_1:\; I^\pi= 0^+,3^-, 4^+, 6^\pm, 7^-,  8^+,  9^\pm,  10^\pm,
             11^-, \;\ldots
                                                                 \label{eqn.08}
\end{equation}
Thus no states with spins $I=1,2$ and 5 belong there. Moreover, one should expect the presence of parity doublets, i.e.,~the strictly symmetric rotor gives the opposite parity states $E_{I^\pm}$ with the spins $I=6,9,10\,\ldots$, at equal energies: $E_{6^-}=E_{6^+}$,  $E_{9^-}=E_{9^+}$, etc. On the other hand, the analog irreducible representations, which can be seen as {\em octahedral} symmetry partners of the sequence in (\ref{eqn.08}), the first of them containing the $I^\pi=0^+$ state are:
\begin{eqnarray}
      A_{1g}:&\;& 0^+,4^+,6^+,8^+,9^+,10^+,\; \ldots\,,\;\;I^\pi=I^+,
                                                                 \label{eqn.09}
                                                                 \\
      A_{2u}:&\;& 3^-,6^-,7^-,9^-,10^-,11^-, \ldots\,,\;I^\pi=I^-,
                                                                 \label{eqn.10}
\end{eqnarray}
according to standard notation. It follows that at the exact tetrahedral symmetry, the lowest tetrahedral band is composed of states with spin-parity combinations given by Eq.\,(\ref{eqn.08}). At the exact octahedral symmetry instead, we should look for the two sequences defined by Eqs.\,(\ref{eqn.09}\,-\ref{eqn.10}). 

\begin{table}
\centering
\caption{$^{152}$Sm: Octahedral deformations, 
         $o_1$ and $o_2$, accompanying an increase in the single tetrahedral
         constraint, $Q_{32}=500,1000,\,\ldots\,2500$\,fm$^3$. Results obtained 
         using the Gogny-HFB method with parameters D1S show that the realistic 
         self-consistent mean-field theory enforces coexistence between
         tetrahedral and octahedral symmetries.
                                                                  \label{tab.01}
         }
\resizebox{\columnwidth}{!}{
\begin{tabular}{cccccccc}
\hline
 \;$t_1=\alpha_{32}$$\phantom{\Bigg|^{}_{}}$ & $o_1=\alpha_{40}$ & $\alpha_{44}$ & $-\alpha_{40}\times\sqrt{\frac{5}{14}}$ & $o_2=\alpha_{60}$ & $\alpha_{64}$ & $\alpha_{60}\times\sqrt{\frac{7}{2}}$ \\ 
\hline
0.0505 & $-$0.0446 & 0.0266 &  0.0267 & 0.0021 & 0.0039 & 0.0039 \\
0.1003 & $-$0.0640 & 0.0383 &  0.0383 & 0.0065 & 0.0122 & 0.0122 \\
0.1483 & $-$0.0850 & 0.0508 &  0.0508 & 0.0152 & 0.0284 & 0.0284 \\ 
0.1930 & $-$0.1117 & 0.0668 &  0.0668 & 0.0288 & 0.0539 & 0.0539 \\ 
0.2343 & $-$0.1402 & 0.0838 &  0.0838 & 0.0460 & 0.0861 & 0.0861 \\
\hline
\end{tabular}
}
\end{table}

One may interpret the results of the total energy calculations in terms of the octahedral symmetry breaking by the tetrahedral one which is the sub-group of the former. Thus one may expect that the resulting rotational level scheme contains traces of {\em both} patterns to some extent and this aspect will be examined below. Table \ref{tab.01} illustrates the co-existence of the tetrahedral and octahedral deformations predicted by our constrained Gogny-Hartree-Fock-Bogoliubov (Gogny-HFB) self-consistent mean-field calculations in agreement with Eqs.\,(\ref{eqn.05}-\ref{eqn.06}); notice the correspondence between deformations in Table \ref{tab.01} and the results for the minima in Fig.\,\ref{fig.02}.

When looking for a set of experimental level-candidates compatible with the group-theory predictions for rotational bands of Eqs. (\ref{eqn.08}) and/or (\ref{eqn.09}\,-\ref{eqn.10}), one is forced by the group theory criteria to apply a number of rather unprecedented selections. To begin with, since high-rank symmetry rotational bands are not expected to generate E2 and E1-transitions, states with the spins and parities given by the above equations but engaging strong transitions of this type must be disregarded in the preselection procedure. On the one hand this condition implies evident experimental difficulties since the spin-parity assignments of such states are particularly difficult and introduces uncertainties. On the other hand following the algorithm presented below provides the clear-cut indications which states should be possibly re-measured.

Next we define a positive-parity {\em reference sequence}, $E_I^{\rm ref}$, composed of states $I^\pi=4^+,6^+,8^+$ and 10$^+$ from Ref.\,\cite{PEG05}. These form a perfect parabola (maximum deviation of 0.6\,per-mil), without E2 connecting-transitions firmly identified, see Ref.\,\cite{MJM13}. Moreover, the lower the spin in this sequence the bigger the number of the associated decay-out transitions, as listed in the third column of Table \ref{tab.02}, showing that this structure is unlikely to resemble any other collective structures in the $^{152}$Sm nucleus. This reference energy sequence defines the reference parabola. In the case of the tetrahedral symmetry this parabola should be common for all the positive and negative parity states given by~Eq.\,(\ref{eqn.08}). We perform a search of all such levels which at the same time are not connected via any E2-transitions within any subsequence of the full sequence in Eq.\,(\ref{eqn.08}). We allowed for weak E2-, and/or E1-, or M1-feeding or 
\begin{table}[h]
\centering
\caption{Experimental energy levels in $^{152}$Sm used in the present analysis. The levels at $I^\pi=4^+$, 6$^+$, 8$^+$, 9$^+$ and 10$^+$ are from Ref.\,\cite{PEG05}. All other levels are from four negative-parity bands in Ref.\,\cite{MJM13}. Columns 3 and 4 give the numbers of decay-out transitions and feeding transitions, respectively. Parentheses are used when given information is uncertain, `none' means that no information exists in the literature. \\}
                                                                 \label{tab.02}
\begin{tabular}{|c|c|c|c|c|}
\hline
Spin      & \;\;E[keV]\;\; &  Nb.D-out              &  Nb.Feed & $\phantom{\Big|}$Reaction$\phantom{\Big|}$ \\[1mm] \hline\hline
3$^-$     &  1579.4        &    10$\phantom{\Big|}$ &   none   &  CE \& $\alpha$   \\[-1mm]
4$^+$     &  1757.0        &    9                   &   1+(1)  &  CE \& $\alpha$   \\
6$^-$     &  1929.9        &    2                   &   (1)    &  CE \& $\alpha$   \\
6$^+$     &  2040.1        &    7                   &   none   &  CE \& $\alpha$   \\
7$^-$     &  2057.5        &    6                   &   2+(1)  &  CE \& $\alpha$   \\
8$^+$     &  2391.7        &    3                   &    1     &  CE \& $\alpha$   \\
9$^-$     &  2388.8        &    4                   &    3     &  CE \& $\alpha$   \\
9$^+$     &  2588\;\;\,    &    2                   &    1     &        $\alpha$   \\
10$^-$    &  2590.7        &    4                   &    1     &        $\alpha$   \\
(10$^+$)  &  2810\;\;\,    &    2                   &   none   &        $\alpha$   \\
11$^-$    &  2808.9        &    2                   &   none   &  CE               \\ \hline
\end{tabular}
\end{table}
decay out of the band but accepted no levels with strong feeding from identified collective-band structures. 

We have verified that all accepted levels, see Table \ref{tab.02}, were essentially populated through $\alpha$-particle reactions or Coulomb excitations (CEs). This latter mechanism is of particular interest here since the states of the discussed band may involve strong E3 and E4 matrix elements and may lead to measurable enhancement of CE cross-sections, a possible independent experimental test.

\begin{figure}[h!]
\begin{center}
\includegraphics[width=0.52\textwidth]{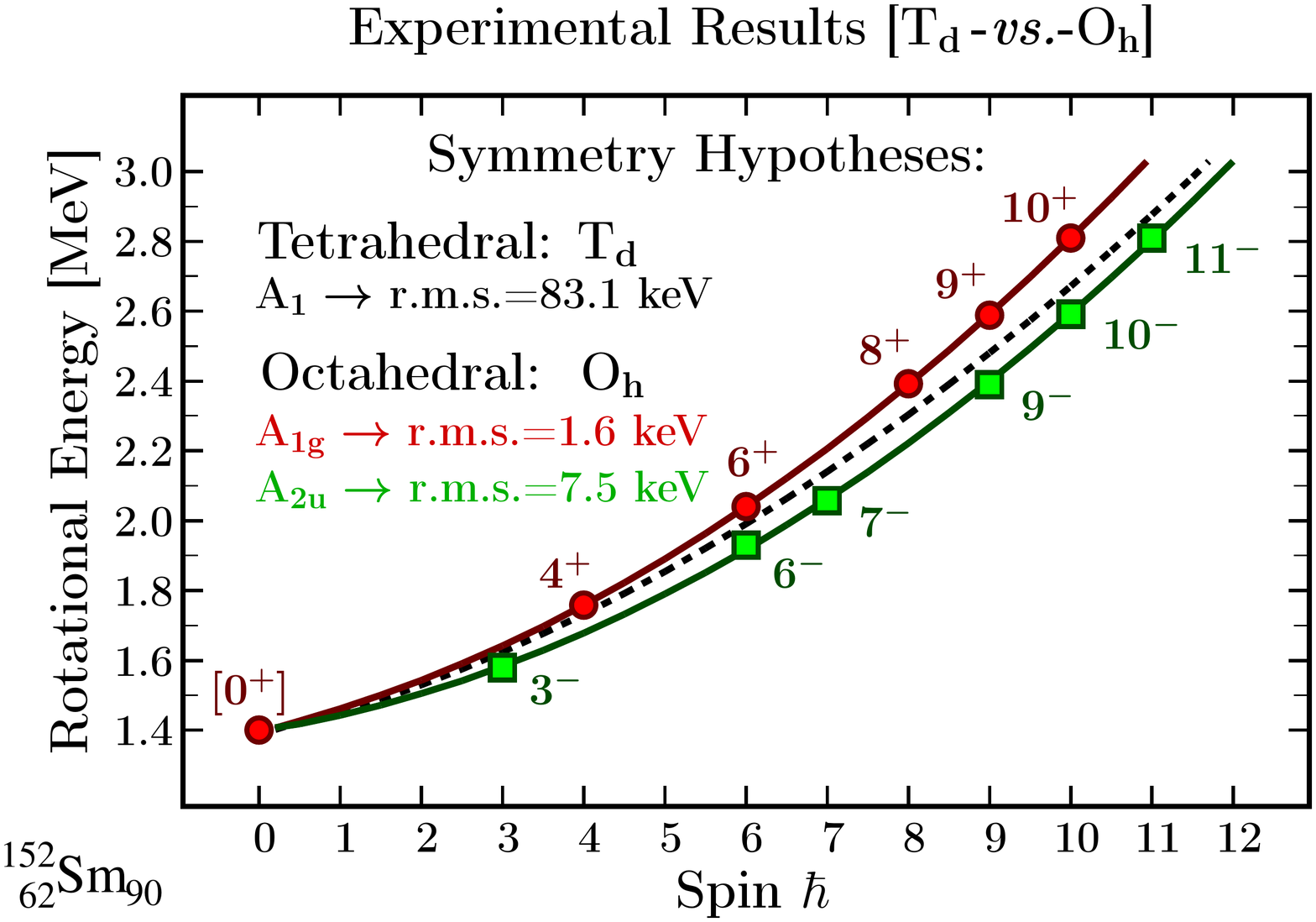}
\end{center}
\vspace{-8mm}
\caption{Graphical representation of the experimental data from Table 
         \ref{tab.02}. Curves represent the fit according to Eq.\,(\ref{eqn.12})
         and are {\em not} meant `to guide the eye'. Point $[0^+]$, marked with 
         square brackets, represents a prediction by extrapolation - it is not 
         an experimental datum.}
                                                                 \label{fig.05}
\end{figure}

To complete the discussion of the results in Table \ref{tab.02}, let us recall the so-called {\em Generalized Quantum Rotor} formalism developed in Ref.\,\cite{JDu01}, cf.~also Ref.\,\cite{AGo08}. It allows to express quantum-rotor Hamiltonians invariant under any given point group symmetry in terms of the expansion onto a tensor-basis, $\{\hat{T}_{\lambda \mu}(\hat{I})\}$, where the nuclear angular momentum is denoted as usual as  
$\hat{I}\equiv\{\hat{I}_+,\hat{I}_-,\hat{I}_0\}$. This formalism as well as our spin-projected Gogny-HFB results obtained exactly as in Ref.\,\cite{STa15} suggest using for the analysis of the high-rank symmetry candidate-bands of Table \ref{tab.02}, the full parabolic expressions of the form 
\begin{equation}
     E_I\approx aI^2+bI+c.
                                                                 \label{eqn.12}
\end{equation} 
We performed the test of the tetrahedral $A_1$-type hypothesis by fitting the parameters of the parabola (\ref{eqn.12}) to the energies in Table~\ref{tab.02}. The root-mean-square deviation between the actual experimental energies and the fitted parabola gives: 
\begin{equation}
   A_1\;\to\;r.m.s.\approx 83.1\,{\rm keV}
   \;\;\leftrightarrow\;\; I^\pi=I^\pm ,
                                                                 \label{eqn.13}
\end{equation}
whereas for the octahedral symmetry hypothesis we have
\begin{eqnarray}
   A_{1g}\;&\to&\;r.m.s.\approx 1.6\,{\rm keV}
   \;\;\leftrightarrow\;\; I^\pi=I^+ ,
                                                                 \label{eqn.14}
                                                                 \\
   A_{2u}\;&\to&\;r.m.s.\approx 7.5\,{\rm keV}
   \;\;\leftrightarrow\;\; I^\pi=I^- .
                                                                 \label{eqn.15}
\end{eqnarray}
Negative parity sequence (\ref{eqn.15}) lies entirely below the positive parity sequence (\ref{eqn.14}), see Fig.\,\ref{fig.05}. Extrapolating the parabolas to zero-spin we find $E^-_{I=0}=1.396\,8$\,MeV compared to $E^+_{I=0}=1.396\,1$\,MeV, a difference of 0.7\,keV at the level 1.4\,MeV excitation. This high precision in the {\em correlation among experimental results} strongly deserves emphasizing. The state $I^\pi=0^+$ in question remains our prediction. We do not find any candidate in the published results -- see however our earlier comments about extra difficulties in populating and detecting this particular state, a remark which, paradoxically, confirms the argumentation. 

We believe that the discussed low r.m.s.~correlations are unlikely an accidental coincidence. However, it will be instructive to assume that the contrary is true and estimate the probability of finding 11 randomly positioned experimental levels close to a single parabola with the r.m.s.$\equiv\delta e$\,$\approx$\,83.1\,keV. We assume that the levels are to be found within an energy stripe defined by the yrast line, $E_I^{\rm yr}$, and an auxiliary line, $E_I^{\rm yr}+\Delta E$. We have verified that the stripe with the width of $\Delta E=1500$\,keV contains all the existing experimental candidate-levels. Since the energy positions are by definition random we assume that the probability of any position is constant within $\Delta E$. Consequently the relative probability for such a random level to be within $\delta e$ is 
$P_{1~\rm level}=\delta e/\Delta E = 83.1/1500\approx 0.055$. Assuming independence of each such `event' from one another we obtain the order of magnitude estimate $P_{11~\rm levels}^{\,\rm r.m.s.=\delta e}=\left(P_{1~\rm level}\right)^{11}\approx 1.1\cdot10^{-14}$. Observe that under similar conditions the probabilities of finding five randomly distributed levels close to a parabola within r.m.s.=1.6\,keV and six randomly distributed levels within r.m.s.=7.5\,keV are  $1.4 \cdot 10^{-15}$ and $1.6 \cdot 10^{-14}$, respectively, fully compatible with the preceding result.


In summary, we have shown that {\em experimental results} for five positive parity levels interpreted as $A_{1g}$ octahedral-symmetry sequence fit a single parabola in agreement with the group-theory interpretation, within over a 1000\,keV interval, with the r.m.s.~deviation of 1.6\,keV. At the same time those for the six {\em experimental}\ negative parity levels interpreted as the $A_{2u}$ octahedral-symmetry sequence fit another single parabola, within the interval of about 1200\,keV, with the r.m.s.~deviation of 7.5\,keV. 

Both reference parabolas extrapolated to $I=0$ meet at an intercept of 1\,396\,keV with 0.7\,keV precision. This suggests that at the low-spin limit the two sequences lie close together resembling approximately the tetrahedral-symmetry band of Eq.\,(\ref{eqn.08}) with the $I^\pi=0^+$ band-head. These observations are compatible with the coexistence of tetrahedral and octahedral symmetries predicted by our calculations and with the group theory criteria. Thus, on the one hand, one may be tempted to conclude that the experimental results illustrated in Fig.\,\ref{fig.05} identify the presence of both discussed symmetries in the $^{152}$Sm nucleus. On the other other hand the data contain uncertainties whose elimination requires specifically designed experiments.

To diminish the experimental uncertainties of the data needed for the analysis illustrated in this work, see Table \ref{tab.02}, one would need a
confirmation of the feeding (see column 4 in the table) in a single type of reaction such as fusion evaporation with $\alpha$ particles and possibly remove uncertainties marked in the Table. Let us emphasize that none of the geometrical nuclear symmetries can be considered exact because of the zero-point motion mechanism within the Bohr collective model and various polarization mechanism e.g.~by particles outside of the high-rank symmetry magic shell closures. Consequently relatively weak electromagnetic transitions are to be expected and this mechanism can or should be used to determine the nature of the weak (`residual') electromagnetic decay. Such experiments would confirm the spin-parity assignments and determine the reduced transition probabilities thus improving the reliability of input data for the presented analysis.

More generally, an unambiguous high-rank symmetry identification procedure for this or any other nucleus must aim at the verification of the group-theory requirement in Eqs.\,(\ref{eqn.08}\ -\ref{eqn.10}) and will very likely follow the patterns presented in this work.


\begin{acknowledgements}
This work was partially supported by the Polish National Science Centre under Contract No.~2016/21/B/ST2/01227 and under the contract No. 2013/08/M/ST2/00257 (LEA-COPIGAL). 
\end{acknowledgements}

\end{document}